\documentclass{aa}
\usepackage{graphicx}
\usepackage{txfonts}
\usepackage{natbib}
\usepackage{longtable}
\usepackage{color}
\usepackage{appendix}

\begin{document}

\def\logg{$\log (g)$~}
\def\Teff{$T_{\rm eff}$~}
\def\FeH{$[Fe/H]$~}
\def\vmicc{$\xi_{t}$}
\def\vmic{$\xi_{t}$~}
\def\msun{$M_{\odot}$~}
\def\gc#1{{\textcolor{blue}{[\bf GC: #1]}}}
\def\cm#1{{\textcolor{red}{[\bf CM: #1]}}}

\title{Barium lines in high-quality spectra of two metal-poor giants in the Galactic halo\thanks{Based on observations made with the UVES at the ESO Very Large Telescope, Paranal Observatory, Chile (ID 098.B-0094(A); P.I. G. Cescutti).}}

\author {G. Cescutti\inst{1,2,3} \thanks {Email: gabriele.cescutti@inaf.it},\,\,\,\,\,
C. Morossi\inst{1},
M. Franchini \inst{1},
P. Di Marcantonio \inst{1}, 
C. Chiappini \inst{4},
M. Steffen \inst{4},
M. Valentini \inst{4},
P. Fran\c cois \inst{5,6},
N. Christlieb\inst{7},
C. Cort\'es \inst{8,9},
C. Kobayashi\inst{10} \& \\
E. Depagne\inst{11}
}

\institute{INAF, Osservatorio Astronomico di Trieste, Via Tiepolo 11,  I-34143 Trieste, Italy
\and
IFPU,  Institute for the Fundamental Physics of the Universe, Via Beirut, 2, I-34151 Grignano, Trieste, Italy
\and
INFN, Sezione di Trieste, Via A. Valerio 2, I-34127 Trieste, Italy
\and
Leibniz-Institut f\"ur Astrophysik Potsdam (AIP), An der Sternwarte 16, 14482, Potsdam, Germany
\and
GEPI - Observatoire de Paris, 64 avenue de l’Observatoire, 75014, Paris, France
\and
Universit\'e de Picardie Jules Verne, 33 rue St-Leu, 80080 Amiens, France
\and
Zentrum f\"ur Astronomie der Universit\"at Heidelberg, Landessternwarte, K\"onigstuhl 12, 69117 Heidelberg, Germany
\and
Departamento de F\'isica, Facultad de Ciencias B\'asicas, Universidad Metropolitana de la Educaci\'on, Av. Jos\'e Pedro Alessandri 774,
7760197 Nu\~noa, Santiago, Chile
\and
Millennium Institute of Astrophysics (MAS), Santiago, Chile
\and
Centre for Astrophysics Research, Department of Physics, Astronomy and Mathematics, University of Hertfordshire, College Lane,Hatfield AL10 9AB, UK
\and
South African Astronomical Observatory (SAAO), Observatory Road Observatory Cape Town, WC 7925, South Africa}
\date{Received xxxx / Accepted xxxx}

\abstract {Theoretical results showed the possibility that neutron capture elements were produced in the early Universe by two different sources: a frequent $s-$process source hosted by rotating massive stars, and a rare $r-$process source hosted most likely by neutron star mergers. The two sources produce barium with different isotopic compositions.}{We aim to investigate the lines of barium in two halo stars, HD~6268 and HD~4306. The spectra present an exquisite quality, both in terms of resolution  ($R>100,000)$ and signal-to-noise ($\sim400$). Due to hyperfine splitting (hfs) effects, barium lines are expected to show slightly different profiles depending on the barium isotopic fraction.}{We applied a standard local thermodynamic equilibrium synthesis of the barium lines. We compared the synthetic results assuming an $s-$process isotopic pattern or an $r-$process isotopic pattern for the two barium lines for each star that exhibited hfs. 
We also applied  a methodology, less dependent on the accuracy of the theoretical Ba hfs structure,  that transforms the lines of HD~4306  into those we would observe if  its  atmospheric parameter values (i.e. $T_{\rm eff}$, log $g$, micro- and macro-turbulence, V$\sin i$, and Ba abundance)  were the same  as those of HD~6268. }{With both methods, our results show that the barium lines with hfs effects of HD~4306 are in agreement with an $s-$process composition and the lines in HD~6268 have a different profile, which is most likely linked to the presence of an $r-$process isotopic pattern.}{Two lines of barium of HD~6268 and HD~4306 seem to confirm the theoretical expectation that both  $r-$process events and also $s-$process contribution by rotating massive stars have polluted the ancient halo of our Galaxy. 
}


\titlerunning{Barium lines in two halo stars}

\authorrunning{Cescutti et al.}

\maketitle

\section{Introduction}

The past decade has seen an explosion of
  elemental abundance determinations for the most metal-poor stars in
  the Galaxy. These new data have shown many surprising results, including the fact that light neutron capture elements (such as Sr, Y, and Zr)
  are more enriched in some extremely metal-poor (EMP) stars compared to the
  expectations of a pure $r-$process pattern \citep{Thielemann11}, and they show an increasing
  scatter towards low metallicities, down to [Fe/H] \footnote{We assumed the common approximation in which iron is a proxy for the entire metal content, and we adopted the notation [A/B]$\equiv$ log$_{10}$(N$_A$/N$_B$)-log$_{10}$(N$_A$/N$_B$)$_{\odot}$ and $\epsilon$(A)$\equiv$ log $_{10}$(N$_A$/N$_H$) for elements A and B.} $\sim -$4 \citep{honda04,BAR05,Franc07,Spite18}.

Astronomers then invoked the need for a new process \citep[e.g.][]{Trava04}
able to provide the additional contribution to light neutron
capture elements. The astrophysical conditions
that would create this additional contribution are still unknown, and
both an $s-$process-like or an r-process-like mechanism was found to reproduce
the abundance pattern between Sr and Ag observed in many of the most
extreme metal poor stars \citep{Montes07}. The s-process production 
in massive stars (weak s-process) was shown to be inefficient 
at low metallicity \citep{Raiteri92}, so most authors have
focused on the existence of a possible second $r-$process
\citep[also called the weak $r-$process; see][]{ArcoMonte11}. 
 
However, the recent nucleosynthesis computations by \citet{Frisch15} showing that rotating massive stars can support the
$s-$process, and hence pollute the early Universe with these elements on
a timescale comparable to that of the $r-$process, have brought a new
twist to the interpretation of the neutron capture element abundances of stars at very low metallicity. 
Interestingly, the existence of rotating massive stars in the
early Universe has already been suggested as the only solution for at
least four 'anomalies' observed in other chemical elements in very
metal-poor stars ([Fe/H] $<-$2.5); namely, the rise of N/O and C/O
towards lower metallicities \citep{Chiappini06}; the low
$^{12}$C/$^{13}$C \citep{Chiappini08} and the primary-like
behaviour of Be and B \citep{Prantzos12}. Recently, we claimed to
have found a fifth signature of the impact of rotating massive stars, namely an early enrichment of the Universe in $s-$process elements \citep{Chiappini11,Cescutti13}. More recently, the impact of rotation in the nucleosynthesis of neutron capture elements was also confirmed by the studies of \citet{LC18} and \citet{Choplin18}. Moreover, chemical evolution models have again confirmed that adopting these yields of rotating massive stars can explain the increase in [Sr/Ba] towards low metallicity \citep{Prantzos18,Rizzuti19,Rizzuti21}. A similar increase was obtained by \citet{Kobayashi20} thanks to the strontium production by electron capture SNe, a potential source of weak r-process; in this case, the barium is formed only by an r-process.  

In this framework, the oldest halo stars can form from an ISM enriched by an $r-$process source or an $s-$process source (or both). A clear prediction of  stochastic models with rotating massive stars is that EMP stars with high [Sr/Ba] ratio should be almost entirely  enriched by the $s-$process \citep[see Fig. 2 in][]{Cescutti14}. On the other hand, an  $r-$process enrichment is expected for stars that show a lower [Sr/Ba] ratio, closer to the $r-$process solar residual\footnote{The residual abundances of chemical species of the solar abundances; after that, theoretical s-process abundances have been removed.} ratio \citep{Arla99}. Following this interpretation, the abundances of the isotopes of the elements produced are expected to present different patterns among these two classes of stars. 
Usually, atomic lines of heavy elements  do not allow us to distinguish the isotopic composition of the element.
However, Ba is an attractive heavy element in this respect. 
In fact, it presents a hyperfine splitting (hfs) of its 4554\,\AA~ line and its 4934\, \AA~ line from the singly ionised stage, which is large enough to be detected \citep{Rutten78} and hence offers the possibility of measuring the fraction of odd isotopes (hereafter, $f_{\rm odd}$\footnote{$f_{\rm odd} = [N(^{135}{\rm Ba}) + N( ^{137}{\rm Ba})]/N({\rm Ba})$}) via resolved asymmetric lines.  According to nucleosynthesis calculations \citep{Arla99}, one expects an $f_{\rm odd}$ = 0.11 $\pm$ 0.01 in the case of a pure $s-$process, and an $f_{\rm odd}$ = 0.46 $\pm$ 0.06 in the case of the $r-$process solar residual. So, in the case of an s-process enrichment, even isotopes  and no hfs effects are almost exclusively expected; on the other hand, in the case of an r-process the impact of the profile of the barium lines is  relatively strong and driven by the large fraction of odd isotopes prone to hfs effects. \citet{Kobayashi20} found a higher $f_{\rm odd}$ = 0.3 for their chemical evolution model with rotating massive stars at [Fe/H]=$-$2.7; this is due to the model, which is in fact a homogeneous one that mixes the contribution of the r-process events and rotating massive stars. The same work, based on recent theoretical computations of nucleosynthesis,  found a slightly higher fraction of odd isotopes ($f_{\rm odd}$ = 0.63) for the pure r-process.

The measurement of the Ba-isotopic ratio, although feasible, is very
challenging.  \citet{Magain95}  measured the isotopic ratio of a very bright halo star, HD 140283, for the first time. He found an $s-$process signature with a high [Sr/Ba]=0.9, which was in agreement with the theoretical expectations of the chemical models of \citet{Cescutti14}.  
However, this result on HD 140283 has been challenged and still needs to
be confirmed \citep[see][]{LambertAllendePrieto02,Collet09,Gallagher10}. 
 \citet{Gallagher12} attempted to measure isotopic ratios in five $s-$process-dominated metal-poor stars, obtaining isotopic fractions more compatible with an s-process-dominated composition.
Indeed, one of the biggest difficulties in this case is properly accounting for 3D effects on the
line formation. \citet{Gallagher15} adopted a 3D modelling  to obtain an r-process isotopic composition for HD 140283, contrary to their previous results in 1D and also differently to the similar attempt in 3D carried out by \citet{Collet09}. In the same year, \citet{Jablonka15} also managed to determine this isotopic fraction in stars belonging to the dwarf spheroidal Sculptor.
 
In this work, we investigated two spectra of extremely high quality (signal-to-noise, $S/N>$400 and $R \sim$ 100,000) for two stars, HD 4306 and HD 6268. The spectra were obtained homogeneously with the same instrument, which was set up and during the same night. 
The final goal is to compare the barium lines of these two stars. For this purpose, the stars were selected to have  parameters as similar as possible: [Fe/H]$\sim-$2.5\,dex, log $g$ $\sim$ 1, and $T_{\rm eff}\sim$ 4500\,K, but
with a different ratio of [Sr/Ba]; so, we expect HD 4306 to show barium with an isotopic pattern typical of an $s-$process, and HD 6268 to show that typical of an $r-$process enrichment. 
The parameters were not as close as ideally desired, but the constraints of bright halo stars with [Fe/H]$<-$2.5 resulted in the choice of these two candidates. The paper is structured as follows: in Sect. 2, we describe the observations and the reductions of the stellar spectra; in Sect. 3, we provide new atmospheric parameters, with a comparison to the previous literature results; in Sect. 4, we present the chemical abundances obtained from equivalent widths; in Sect. 5, we show the comparison of the synthetic spectra with $r-$ and $s-$process isotopic compositions to the observational ones; in Sect. 6, a new empirical method is described to directly compare the two observational spectra; in Sect. 7, we draw our conclusions.

\section {Observations and data reduction}

The observations were performed using the UVES high-resolution
spectrograph \citep{Dekker00}, mounted at the UT2 Kueyen
Telescope at the ESO Paranal Observatory (Chile), in slit mode.  We adopted the
standard R530 set-up (red arm only, cross disperser 3,  centred at
5200\, covering the 4140--6210\,\AA~ wavelength interval with a
resolving power of $R~\sim$100,000.

  The spectra were acquired on the night of 18th October 2016,
with three exposures  of 2400 seconds (40 minutes) for HD\,4306 and two exposures of 1600 seconds (26 minutes and 40 seconds) for HD\,6268.  Each of the resulting spectra have approximately  
  an average S/N of 250 for HD 4306 and 350 for HD 6268. Details of the observations are summarised in Table~\ref{tablog}.

\begin{table*}[ht!]
\begin{minipage}{180mm}
\caption{Observing log}

\begin{tabular}{|c|c|c|c|c|c|c|c|c|}
  \hline
  Object  & R.A.                 & Dec. & B & G & Exp. & $<S/N>$ & Obs. Date &  No of   \\
             &  (J2000.0)         & (J2000.0) & (mag) & (mag) & time (s) & &   (UT)  & exp.s   \\
                                                                 
 \hline
 
HD 4306 &  00 45 27  & -09 32 44 & 9.71 & 8.76 &  2400s & 250 &19th Oct. 2016 & 3 \\

\hline

HD 6268 & 01 03 18    & -27 52 54  & 8.89 & 7.80 &  1600s & 350 &19th Oct. 2016 & 2 \\

\hline

\end{tabular}
\label{tablog}
\end{minipage}
\end{table*}

The spectra were reduced using the standard UVES pipeline with the following exceptions:
\begin{itemize}
\item 23 high-quality flat fields were purposely obtained and used to build the MASTERFLAT;
\item  wavelength calibration was achieved by using ThAr lamps obtained immediately after the stellar observations; 
\item flat-fielding is obtained in pixel-pixel space before extraction (reduce.ffmethod=pixel) and not using the default option 'extract';
\item all the extracted spectral orders were normalised individually before merging.
\end{itemize}

Then, the spectra of each star were Doppler$-$shifted to put them in the laboratory
reference frame, cleaned by cosmic ray, and, eventually averaged to achieve the highest possible $S/N$. Using DER\_SNR, a 'simple and general spectroscopic
signal-to-noise measurement algorithm' \citep{Stoehr07}, we obtained S/N $\simeq$ 390/640 and S/N $\simeq 430/720$, for HD\,4306 and HD\,6268 respectively, in the
blue and in the red parts of the spectra.

\section {Atmospheric parameters}
We re-determined the stellar parameters of HD \,4306 and HD \,6268
in order to take advantage of the very high S/N of our mean spectra, even if these stars have been studied several times in the literature; among the most recent results with a detailed chemistry, we used the works by \citet{honda04}, \citet{Roederer14}, and \citet{mashonkina17} for a direct comparison. The full list of  sixteen original  atmospheric parameter determinations for HD 6268 is presented in the appendix (Table \ref{lit6268}), with the first determination from 40 years ago by \citet{1981ApJ...244..919L}, and the most recent one by \citet{2019AA...627A.138A}.

\begin{figure}[ht!]
\begin{minipage}{90mm}

\includegraphics[width=85mm]{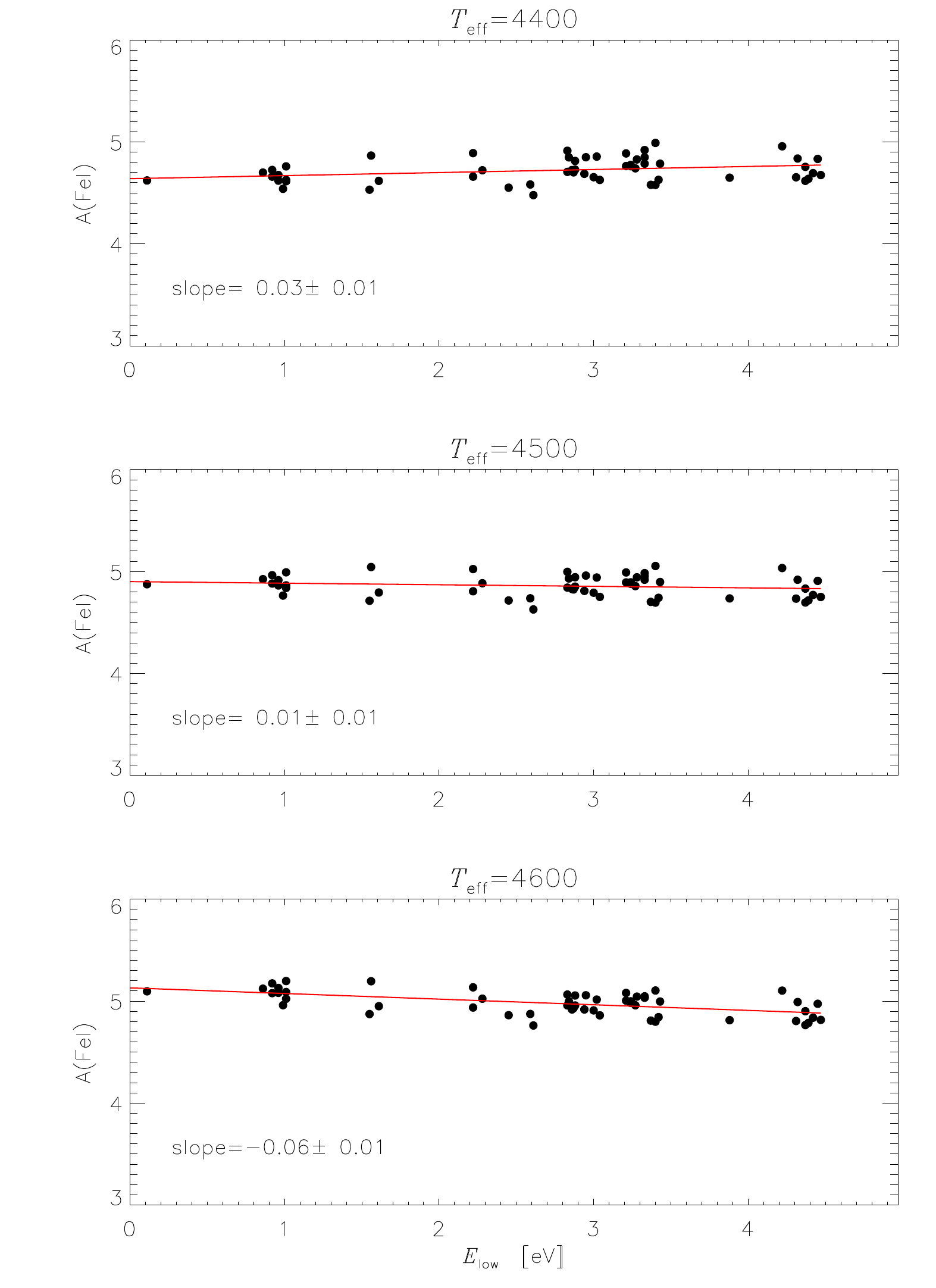}
\caption{ Iron abundance from Fe I lines versus excitation potential of HD\,6268 for the adopted $T_{\rm eff}$=4500\,K (middle panel)  and for a temperature lower ($T_{\rm eff}$=4400\,K, top panel) and higher ($T_{\rm eff}$=4600\,K, bottom panel).}\label{TeffElow}
\end{minipage}
\end{figure}

We derived the stellar parameters of the two stars via
the spectroscopic approach; namely, the effective temperature
$T_{\rm eff}$ was determined by requiring the iron abundance from Fe I lines to be  independent of their excitation potential, and the surface
gravity log\,{\it g} was derived from the ionisation equilibrium of
Fe\,I and Fe\,II . The microturbulence velocity $\xi$ was estimated
by requiring the [Fe/H] derived from the Fe I lines to be
independent of their equivalent widths. 
 The equivalent widths of  a list of 
81~Fe\,I and of 14~Fe\,II lines and of 99~Fe\,I and 16~Fe\,II lines were measured in the spectra of HD\,6268 and HD\,4306, respectively.
Then, an iterative procedure was adopted
to derive the stellar parameters using as initial guess for the  parameter values those
taken from the latest literature. We used the ABUNDANCE and the BLACKWEL codes\footnote{http://www.appstate.edu/~grayro/spectrum/spectrum.html} to derive estimates of $T_{\rm eff}$, log~$g$, [Fe/H], and  $\xi$. 
 In Fig. \,\ref{TeffElow}, we show that at the adopted temperature  $T_{\rm eff}$=4500\,K the iron abundance of HD\,6268 obtained from Fe I lines  is almost independent of their excitation potential; on the other hand, a significant slope is obtained at a temperature  lower/higher of 100\,K.
The atmosphere models of the two stars were calculated for the newly derived stellar parameters by using ATLAS12 \citep{kurucz05}. As discussed in \citet{kurucz05} and \cite{castelli05},
ATLAS12 can generate an atmosphere model for any desired
 chemical composition of individual elements and microturbulent velocity, since its treatment of opacity is based on the
opacity sampling technique instead of the use of opacity distribution
functions as, for example, in ATLAS9.

\begin{table*}[ht!]
\caption{Atmospheric parameters \label{tab:par}}
\begin{tabular}{|c|c|c|c|c|}
\hline
 \multicolumn{5}{|c|}{HD\,4306} \\
 \hline
$T_{\rm eff}$ & log~$g$ & [M/H] & $\xi$ & Reference \\
K & dex & dex & km~s$^{-1}$& \\ 
 \hline
4700$\pm$ 50 &  1.80$\pm$ 0.10 &   -2.90$\pm$ 0.10 &   1.50$\pm$ 0.13 &this paper \\  
\hline
 4810 &  1.80 &   -2.89 &   1.60 & \citet{honda04} \\
 4960 & 2.18 & -2.74 & 1.30 & \citet{mashonkina17} \\
 \hline 
 \multicolumn{5}{|c|}{HD\,6268} \\
 \hline
$T_{\rm eff}$  & log~$g$ & [M/H] & $\xi$ & Reference \\
K & dex & dex & km~s$^{-1}$ & \\
\hline
4500$\pm$ 50 &   0.80$\pm$ 0.10 &  -2.50$\pm$ 0.10 & 2.25$\pm$ 0.13 & this paper \\
\hline
4600 &   1.00 &  -2.63 & 2.1 &\citet{honda04}\\
 4570 & 0.70 & -2.69 & 1.85 & \citet{Roederer14}\\
\hline

\end{tabular}
\end{table*}
Table~\ref{tab:par} shows the obtained atmospheric parameters for HD\,4306 and HD\,6268 together with estimates from the literature. 
We found atmospheric parameters that are reasonably close with those available in literature, even if we estimate lower effective temperatures for both stars; in fact, both  \citet{honda04} and \citet{mashonkina17} estimated $T_{\rm eff}$ with a different approach based on the photometric data. On the contrary, in \citet{Roederer14} $T_{\rm eff}$ is derived with our same approach, by requiring that abundances derived from Fe\,I lines show no trend with the excitation potential and indeed their estimated $T_{\rm eff}$ for HD\,6268 is the closest to our result.


\section{Chemical abundances from equivalent widths}

We used the measured equivalent widths of other weak lines present in our spectra to derive abundances for the elements listed in Table\,\ref{tab:ele}.  The errors in our abundances reported in the table are based only on the standard deviation to the mean from the abundances measured in the different lines. These errors are considered as lower limits because we did not compute the impact of the atmospheric parameters on the error budget for all the elements. Nevertheless, in Table \ref{error_atm} we show the impact of the variation of the  atmospheric parameters on the abundance of barium for HD 6268. The variation is small, 0.03 dex, and there is a total budget of 0.08 dex for the errors on the abundance of barium of this star; a similar impact is expected for HD 4306.
For Sr, we can measure only one  line, and for this reason we cannot compute this error. 
It is worth noticing that after deriving the abundances of the different elements with the code ABUNDANCE, we used them to recompute the corresponding atmosphere models with ATLAS12 for the two
stars, and we iterated in order to have a full consistency between the abundances derived from the spectra and those used in constructing the atmosphere structures.

\begin{table*}[ht!]
\caption{Chemical abundances of the two stars and comparison to similar results in literature. For solar abundances - [X/H] - we considered \citet{Grevesse2007}. For the literature data, we also provide the information on how the the abundances were computed assuming the LTE approximation or NLTE.}\label{tab:ele}
\begin{tabular}{|c|c|c|c|c|c|c|c|}
\hline
\multicolumn{8}{|c|}{}  \\
\multicolumn{8}{|c|}{HD\,4306}  \\
\hline
\multicolumn{8}{|c|}{}  \\
\multicolumn{4}{|c|}{this paper} &  \citet{honda04} & \multicolumn{2}{|c|}{\citet{mashonkina17}}&\citet{Roederer14}\\
Elem  & [X/H] - LTE & A(X) &N lines & A(X) - LTE & A(X) LTE & A(X) NLTE & A(X) -LTE\\
&&&&&&&\\
Mg I & -2.49$\pm$0.09 & 5.04 & 4 &  5.25 &  5.27 & 5.30 & - \\
Ca I & -2.49$\pm$0.02 & 3.82 &14 &  3.97 &  3.98 & 4.11 & - \\
Sc II & -2.75$\pm$0.07& 0.42 & 5 &  0.45 &  -    & -    & - \\
Ti I & -2.77$\pm$0.02&  2.13 &11 &  2.55 &  2.41 & 2.63 &-  \\
Ti II & -2.33$\pm$0.04& 2.57 &23 &  2.55 &  2.58 & 2.59 & - \\
Cr II & -2.74$\pm$0.04& 2.90 &2  &  2.97 &  -    & -    & - \\
Mn I & -3.67$\pm$0.01&  1.72 & 2 &  2.08 & -     & -    & - \\
Fe I & -3.00$\pm$0.01&  4.45 &99 &  4.62 &  4.60 & 4.71 & - \\
Fe II & -2.90$\pm$0.02& 4.55 &16 &  4.62 &  4.76 & 4.76 & - \\
Sr II & -3.27         &-0.35 & 1 & -0.08 &  0.17 & 0.13 & - \\
Ba II & -4.05$\pm$0.07&-1.88 & 2 & -1.84 & -1.62 &-1.62 & - \\
&&&&&&&\\
\hline
\multicolumn{8}{|c|}{}  \\
\multicolumn{8}{|c|}{HD\,6268}\\
\hline
\multicolumn{8}{|c|}{}  \\
\multicolumn{4}{|c|}{this paper} &  \citet{honda04} & \multicolumn{2}{|c|}{\citet{mashonkina17}}&\citet{Roederer14}\\
Elem  & [X/H] - LTE & A(X) &N lines & A(X) - LTE & A(X) LTE & A(X) NLTE & [X/H] -LTE\\
&&&&&&&\\
Mg I & -2.13$\pm$0.17   & 5.40 &  4 &  5.44 & - & - &  5.35 \\
Ca I & -2.27$\pm$0.02   & 4.04 & 13 &  4.09 & - & - &  3.98 \\
Sc II & -2.65$\pm$0.08  & 0.52 &  4 &  0.51 & - & - &  0.39 \\
Ti I & -2.55$\pm$0.02   & 2.35 & 10 &  2.62 & - & - &  2.25 \\
Ti II & -2.23$\pm$0.04  & 2.67 & 14 &  2.62 & - & - &  2.54 \\
Cr II & -2.37$\pm$0.11  & 3.27 &  2 &  3.20 & - & - &  3.11\\
Mn I & -3.10$\pm$0.01   & 2.29 &  2 &  2.49 & - & - &  2.54\\
Fe I & -2.51$\pm$0.01   & 4.94 & 81 &  4.88 & - & - &  4.68 \\
Fe II & -2.51$\pm$0.03  & 4.94 & 14 &  4.88 & - & - &  4.81 \\
Sr II & -2.59           & 0.33 &  1 &  0.36 & - & - &  0.33 \\
Ba II & -2.62$\pm$0.07  &-0.45 &  2 & -0.45 & - & - & -0.46 \\
&&&&&&&\\
\hline

\end{tabular}
\end{table*}

\begin{table*}[ht!]
\caption{Variations of barium abundances obtained from the lines at 5853.668\AA and 6141.730 \AA  \, as a function of the atmospheric parameters. We present the abundance derived from the single barium lines with three digits; in fact, they are considered to compute the mean and $\Delta$ [Ba/H] (the following as usual with two digits).}  \label{error_atm}
\begin{tabular}{|c|c|c|c|c|c|}
\hline
     &  best model &    $\Delta$ \Teff=50K &$\Delta$ \Teff=$-$50K  &$\Delta$ \logg=+0.1 &$\Delta$ \logg=$-$0.1 \\
\hline
[Ba/H] from 5853.668 \AA & -2.611  & -2.581  & -2.639 & -2.585 & -2.637\\

[Ba/H] from 6141.730 \AA & -2.635  & -2.599  & -2.669 & -2.608 & -2.662\\
mean [Ba/H]                  & -2.62   & -2.59   & -2.65  & -2.60  & -2.65  \\
\hline
$\Delta$ [Ba/H] &                &  0.03   &  -0.03 &   0.02   & 0.03  \\
\hline
\end{tabular}
\end{table*}

Even if a detailed study of the chemical composition of HD\,4306 and HD\,6268 is not the purpose of this paper, in which we want to study the profiles of Ba\,II lines, the results shown in Table\,\ref{tab:ele} indicate that, starting from our measured equivalent widths and adopting the derived atmosphere parameters, we obtain abundances that are consistent with those reported in literature. We thus provide further confirmation of the reliability of the results reported in  Table\,\ref{tab:par}. Regarding the abundances, the enrichment in terms of neutron capture elements in the two stars is quite different: 
HD 6268 is more abundant in heavy elements compared to HD 4306 and has a ratio of [Sr/Ba]$\sim$0; with abundances of [Sr/Fe]$<$0 and [Ba/Fe]$<-$1, HD 4306 presents a high [Sr/Ba]$>0.75$. 
It is important to confirm this enrichment pattern, since it is on this basis that we select these stars to have, respectively, an expected barium with r- or s- process isotopic ratios, following the theoretical outcomes of \citet{Cescutti14}.
It must be pointed out that the different abundance values we derived for Fe\,I and Fe\,II and for Ti\,I and Ti\,II are the result of deviations from local thermodynamic equilibrium (LTE), in particular
for Fe\,I and Ti\,I \citep[see][]{mashonkina17}. It is also worth underlining that the impacts of the non-local thermodynamic equilibrium (NLTE) approach to the barium abundances are negligible for HD 4306 according to the results in \citet{mashonkina17}. On the other hand, 
the following estimates were also obtained for the star HD 4306 (\Teff = 4700; \logg = 1.8; [Fe/H] = -2.8): for the barium abundance of -1.88, $\delta$ NLTE = +0.14 for the 6141 line and $\delta$NLTE = +0.21 for the 4554 line; for the barium abundance of -1.62, $\delta$NLTE = +0.11 for the 6141 line and $\delta$NLTE = +0.17 for the 4554 line; for the barium abundance of -1.32, $\delta$NLTE = +0.09 for the 6141 line and $\delta$NLTE = +0.09 for the 4554 line \citep[][and Sergey Korotin private communication ]{Korotin15}.  With these estimates of the NLTE impact, we assume an abundance for Ba too high by 0.06-0.07 dex by considering the abundance of line at 6141\AA~ for the line at 4554\AA. Indeed, the difference is not negligible, but still within the error. Finally,  we also expect NLTE to have a very small impact on Ba for HD 6268, following the results for a similar star in terms of stellar atmosphere and barium abundance again in the \citet{mashonkina17} dataset  (UMi 36886).

\begin{figure*}[ht!]
\begin{minipage}{180mm}

\includegraphics[width=180mm]{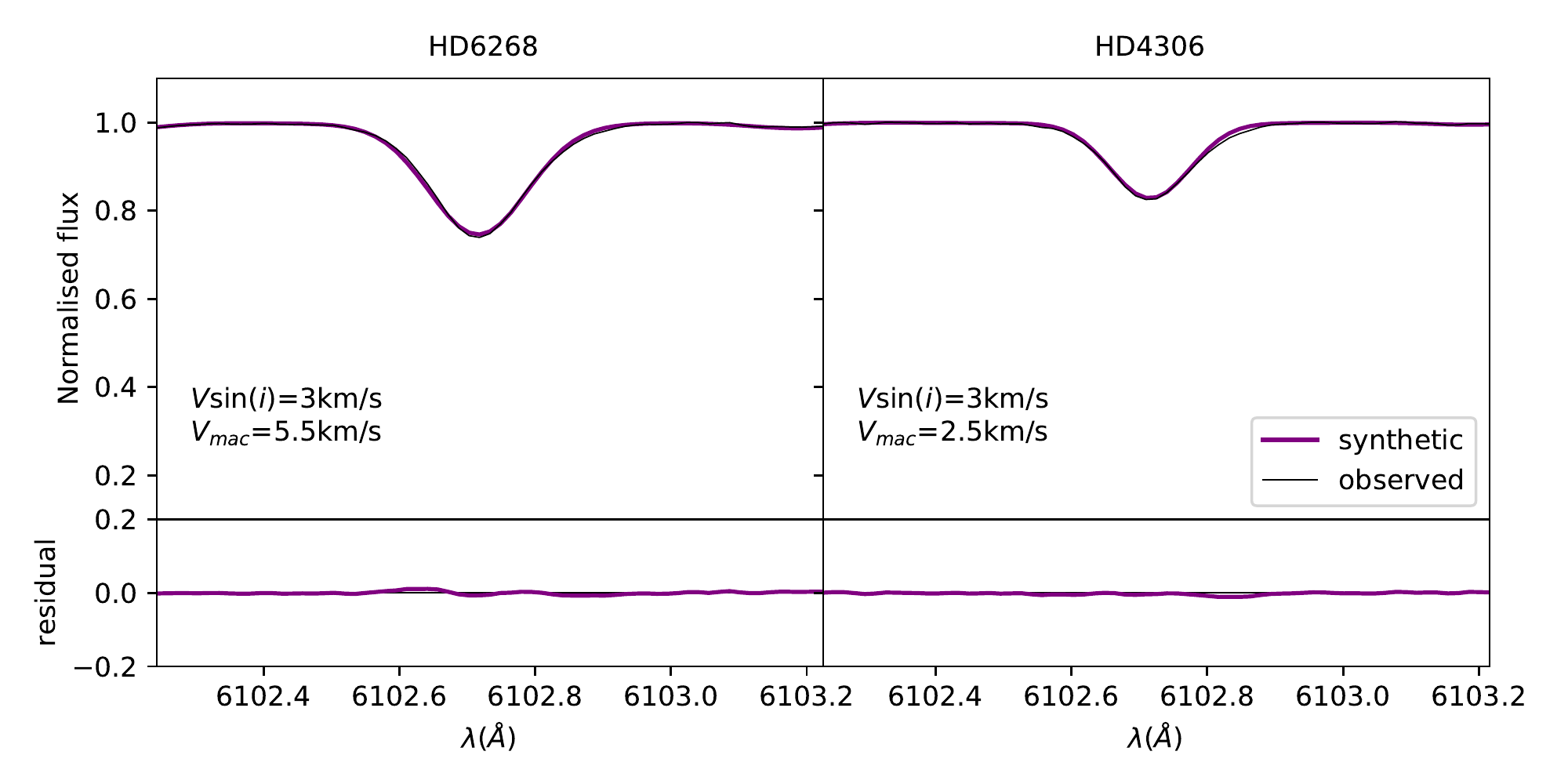}

\caption{Ca\,I line at 6102.7~\,\AA. In the upper panel, we show the observed spectra with a black line. The purple lines are the synthetic spectra obtained; in the lower panel, the residuals between synthetic spectra and observed spectra are reported. In the left panel, we present the results for HD\,6268, and on the right we show those for HD\,4306.}\label{f6101_ca}

\end{minipage}
\end{figure*}

\begin{figure*}[ht!]
\begin{minipage}{180mm}

\includegraphics[width=180mm]{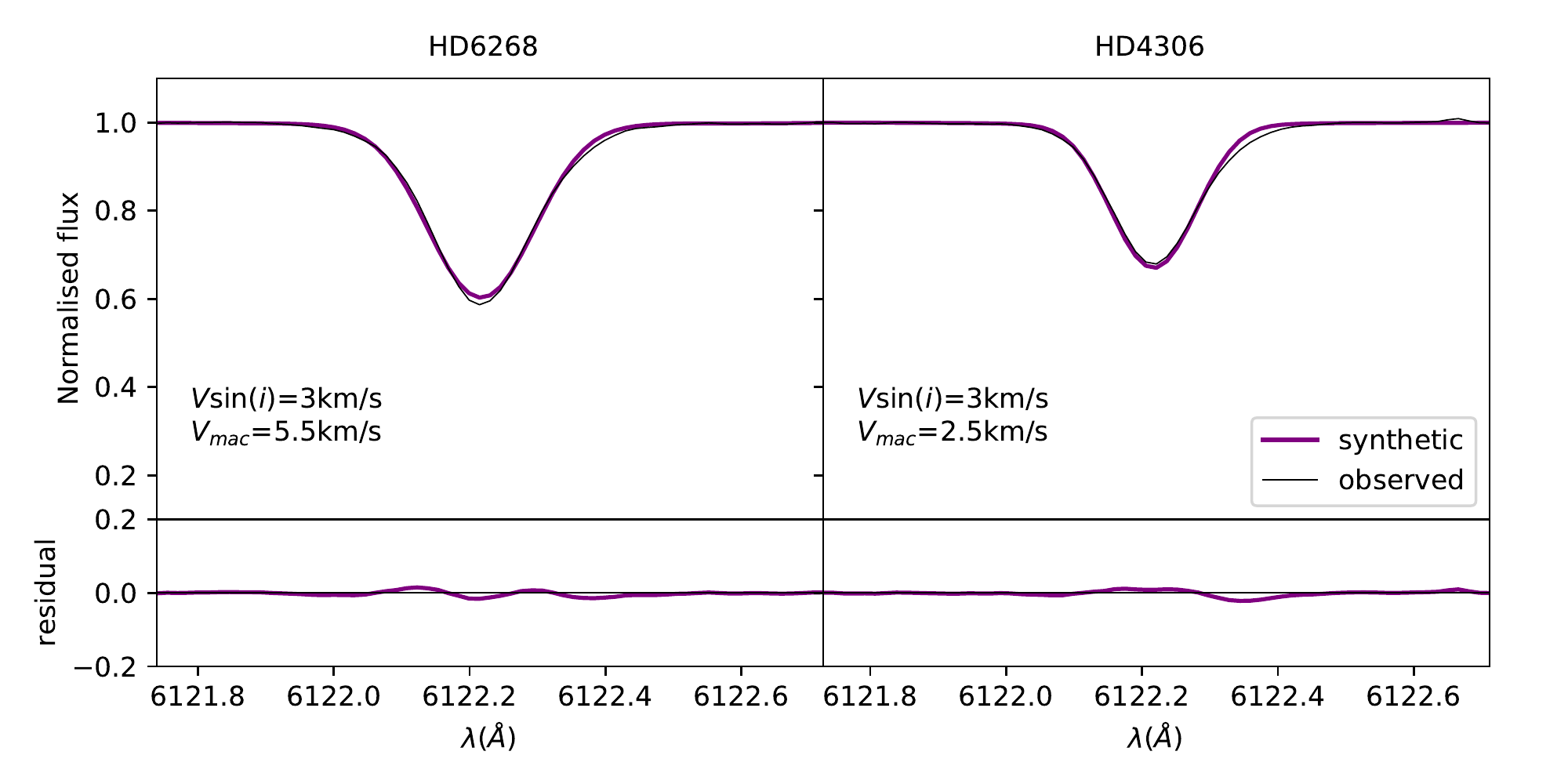}

\caption{Same as Fig. \ref{f6101_ca},  but for Ca\,I line at 6122.2~\,\AA.}\label{f6122_ca}

\end{minipage}
\end{figure*} 

\begin{figure*}[ht!]
\begin{minipage}{180mm}

\includegraphics[width=180mm]{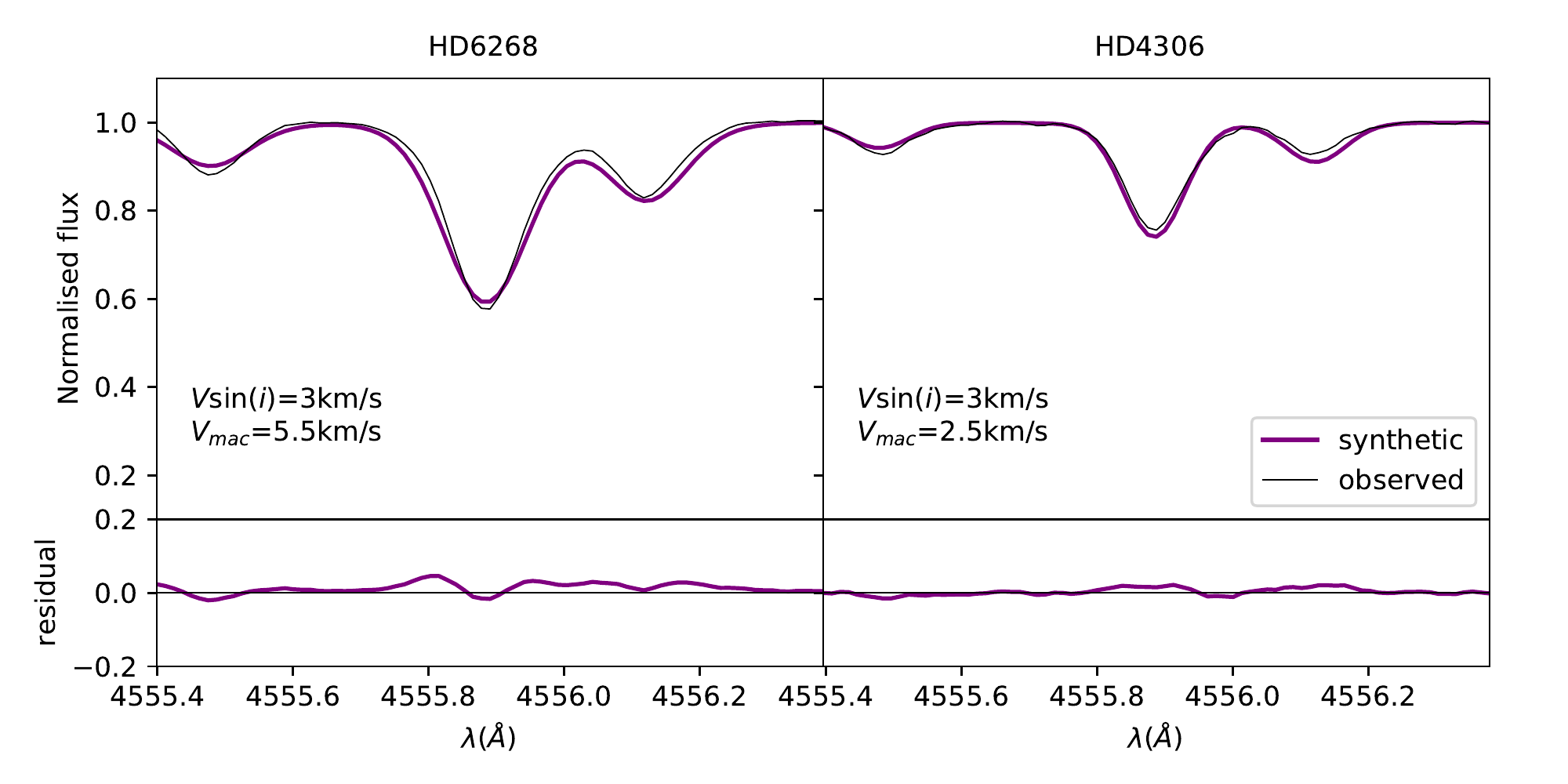}

\caption{Same as Fig. \ref{f6101_ca}, but for Fe\,II line at 4555.8~\,\AA.}\label{f4555_fe}

\end{minipage}
\end{figure*}

\section{Spectral Synthesis} \label{subsec:synth}
\subsection{Synthesis of calcium and iron lines}

Using the obtained atmosphere models and the stellar abundances, we computed the LTE synthetic spectra of HD\,4306 and of HD\,6268 with SPECTRUM. In order to compare them with those observed, we need to take into account the broadening introduced by the instrument and by macroturbulence, $V_{\rm mac}$, and rotation, $V\sin i$. The shape of the instrumental profile used  in degrading the synthetic spectra to the UVES resolution was derived by fitting the profile of several emission lines in the ThAr lamp spectra taken during the observing run. This kind of analysis provides an instrumental profile, which is a linear combination of a Gaussian and a Lorentzian both with $R$=110,000 and with relative weights of 0.9 and 0.1, respectively. As far as macroturbulence and rotation broadening are concerned, we compared the observed and broadened synthetic profiles of three lines: CaI at 6102.7\,\AA, CaI at 6122.2\,\AA,~ and FeII at 4555.8\,\AA. We used the combination of Gaussian profiles computed with $V_{\rm mac}$ varying from 0 to 6\,km~s$^{-1}$ at step 0.5\,km~s$^{-1}$ and rotational profiles \footnote{See Eq.\,18.14 in \citet{gray08}; that is, flux convolution approximation including the limb-darkening coefficient $\epsilon$ (in this work $\epsilon$=0.6).} with $V\sin i$=0,1,2,3, and 4\,km~s$^{-1}$ as broadening functions. In Figs.\,\ref{f6101_ca}, \ref{f6122_ca}, and \ref{f4555_fe}, we show the best resulting synthetic broadened profiles compared to the observed ones, achieved by using $V_{\rm mac}$=2.5 and 5.5\,km~s$^{-1}$ and  $V\sin i$=3.0 and 3.0\,km~s$^{-1}$ for HD\,4306 and HD\,6268, respectively. 

\subsection{Synthesis of barium lines}

The same broadening was then adopted to obtain synthetic line profiles for four Ba\,II lines, namely the lines at  4554.0, 4934.1, 5853.7, and 6141.7\,\AA.
The first two lines' profiles are strongly dependent on the Ba isotope ratios, while the other two are almost independent of them. In computing the theoretical line profiles, we used the $s-$ and $r-$ process contributions to the five isotopes from \citet{arlandini99} to construct pure $s-$process and pure $r-$process Ba~II lines. The Ba hfs line list is the one we used for the resonant lines at 4554 \AA line and 4934 \AA  is from \citet[][Table~1]{mcwilliam98}, and it is  shown here in Table \ref{hyperfine} together with the mass fractions considered for each isotope in the $s-$process and $r-$process cases.
\begin{table*}
\caption{Parameters of barium lines at 4554 \AA and 4934 \AA\,used in our calculations with hfs, including the wavelengths, excitation potentials of the lower level, numbers of isotopes, the mass fraction for each isotope, and oscillator strengths.}\label{hyperfine}
\begin{center}
\begin{tabular}{|l|c|r||c|r|c|c|}
\hline
 \multicolumn{1}{|c|}{Ba II} & \multicolumn{2}{|c||}{E$_{low}$=0.00 eV  log gf=+0.17}&  \multicolumn{2}{|c|}{E$_{low}$=0.00 eV  log gf=$-$0.15}&$s-$process&$r-$process\\
& Wavelength (\AA) & Strength & Wavelength (\AA) & Strength &fraction&fraction\\
\hline
$^{134}Ba$ & & &&&0.02&0.00\\
&  4554.032 & 1.000 & 4934.075 & 1.000&&\\
\hline
$^{135}$Ba &&&&&0.03&0.40\\
&  4554.001 & 0.1562& 4934.034 & 0.3125&&\\
&  4554.002 & 0.1562& 4934.045 & 0.0625&&\\
&  4554.003 & 0.0625& 4934.093 & 0.3125&&\\
&  4554.049 & 0.4375& 4934.104 & 0.3125&&\\
&  4554.052 & 0.1562&&&&\\
&  4554.053 & 0.0313&&&&\\
\hline
$^{136}$Ba & &&&&0.10&0.00\\ 
&4554.032 & 1.000 & 4934.075 & 1.000 &&\\
\hline
$^{137}$Ba &&&&&0.09&0.32 \\
&  4553.997 & 0.1562 & 4934.029 & 0.3125&&\\
&  4553.999 & 0.1562 & 4934.041 & 0.0625&&\\
&  4554.000 & 0.0625 & 4934.096 & 0.3125&&\\
&  4554.052 & 0.4375 & 4934.107 & 0.3125&&\\
&  4554.054 & 0.1562&&&&\\
&  4554.055 & 0.0313&&&&\\
\hline
$^{138}$Ba &&&&&0.76&0.28 \\
&  4554.032 & 1.0000 & 4934.075 & 1.0&&\\
\hline
\end{tabular}
\end{center}
\end{table*}

\begin{figure*}[ht!]
\begin{minipage}{180mm}

\includegraphics[width=180mm]{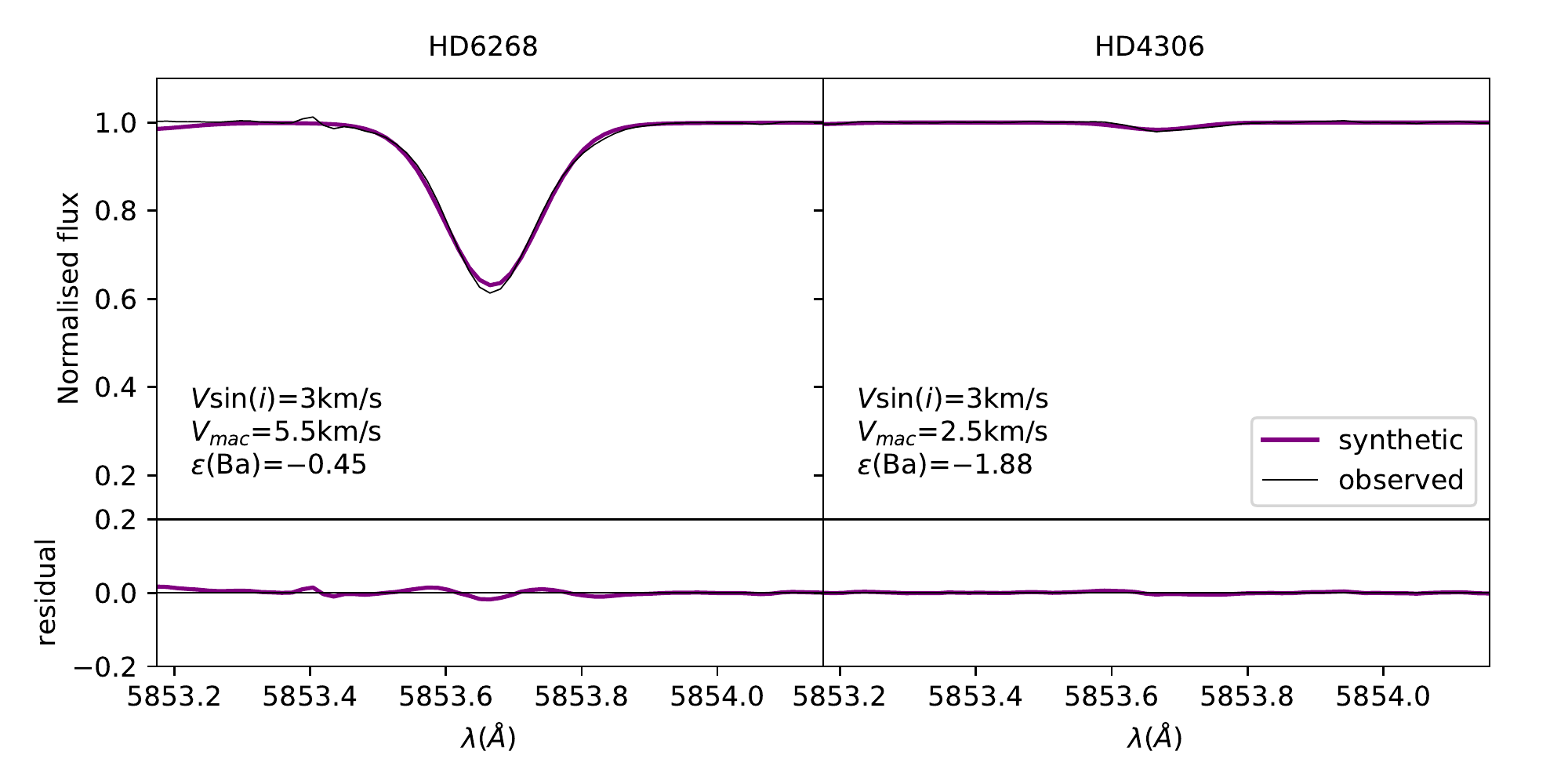}

\caption{Ba II line at 5853\,\AA. In the upper panel, we show the observed spectra with a black line. The purple line is actually two nearly identical profiles, computed for pure s- and pure r-process hfs, respectively; in the lower panel, the residuals between synthetic spectra and observed spectra are reported. In the left panel, we present the results for HD\,6268, and on the right we show those for HD\,4306.}\label{f5853}

\end{minipage}
\end{figure*}

\begin{figure*}[ht!]
\begin{minipage}{180mm}

\includegraphics[width=180mm]{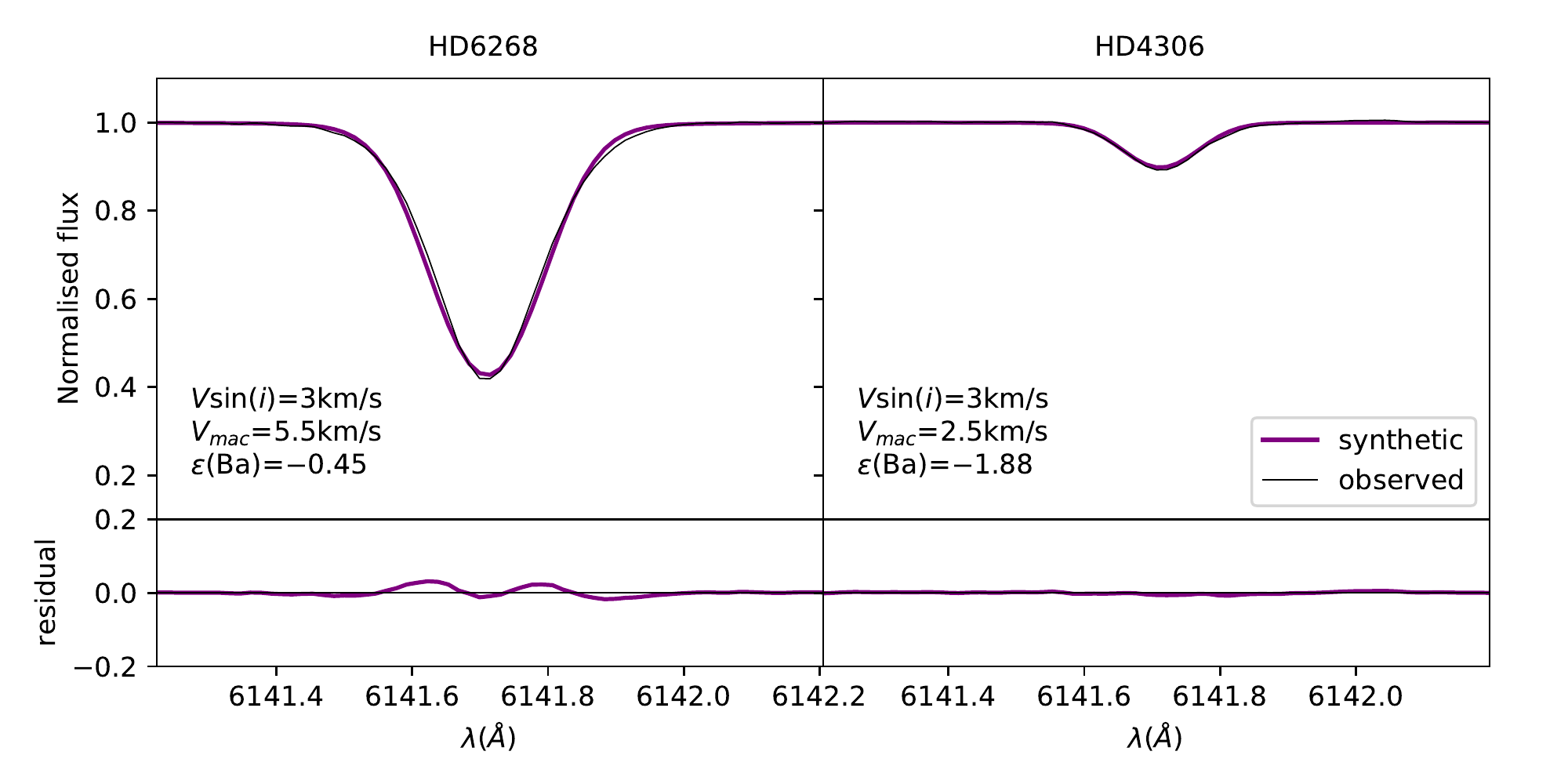}

\caption{Same as Fig. \ref{f5853}, but for Ba II line at 6141\,\AA.}\label{f6141}

\end{minipage}
\end{figure*}

\begin{figure*}[ht!]
\begin{minipage}{180mm}

\includegraphics[width=180mm]{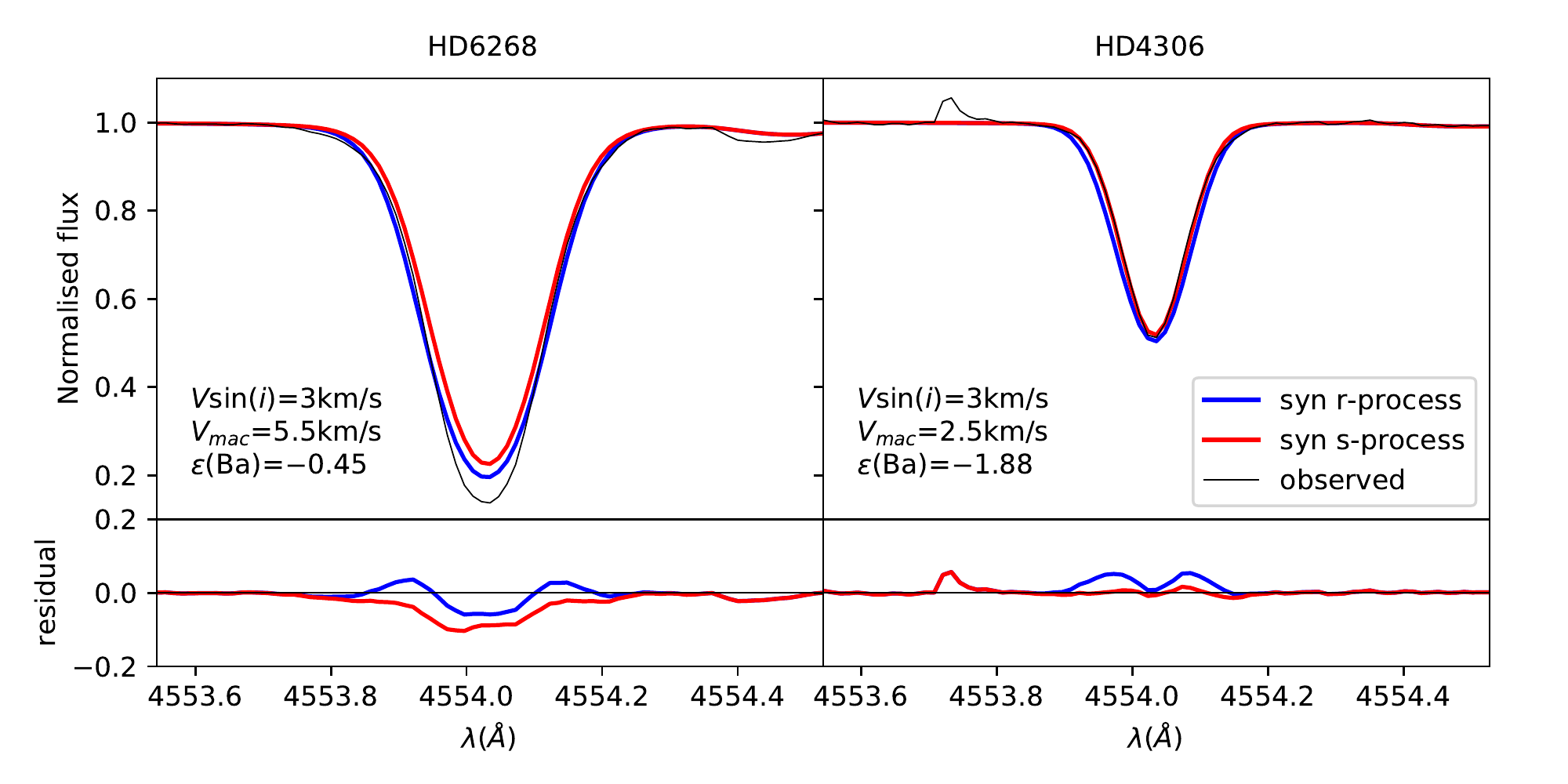}

\caption{Ba II line at 4554\,\AA. In the upper panel, we show the observed spectra with a black line.  The blue lines are the synthetic spectra obtained with  an $r-$process isotopic composition of barium. The red lines are the synthetic spectra obtained with an $s-$process composition. In the lower panel, the residual for both synthetic spectra are reported. In the left panel, we present the results for HD\,6268, and
on the right we show those for HD\,4306.}\label{f4554}

\end{minipage}
\end{figure*}

\begin{figure*}[ht!]
\begin{minipage}{180mm}

\includegraphics[width=180mm]{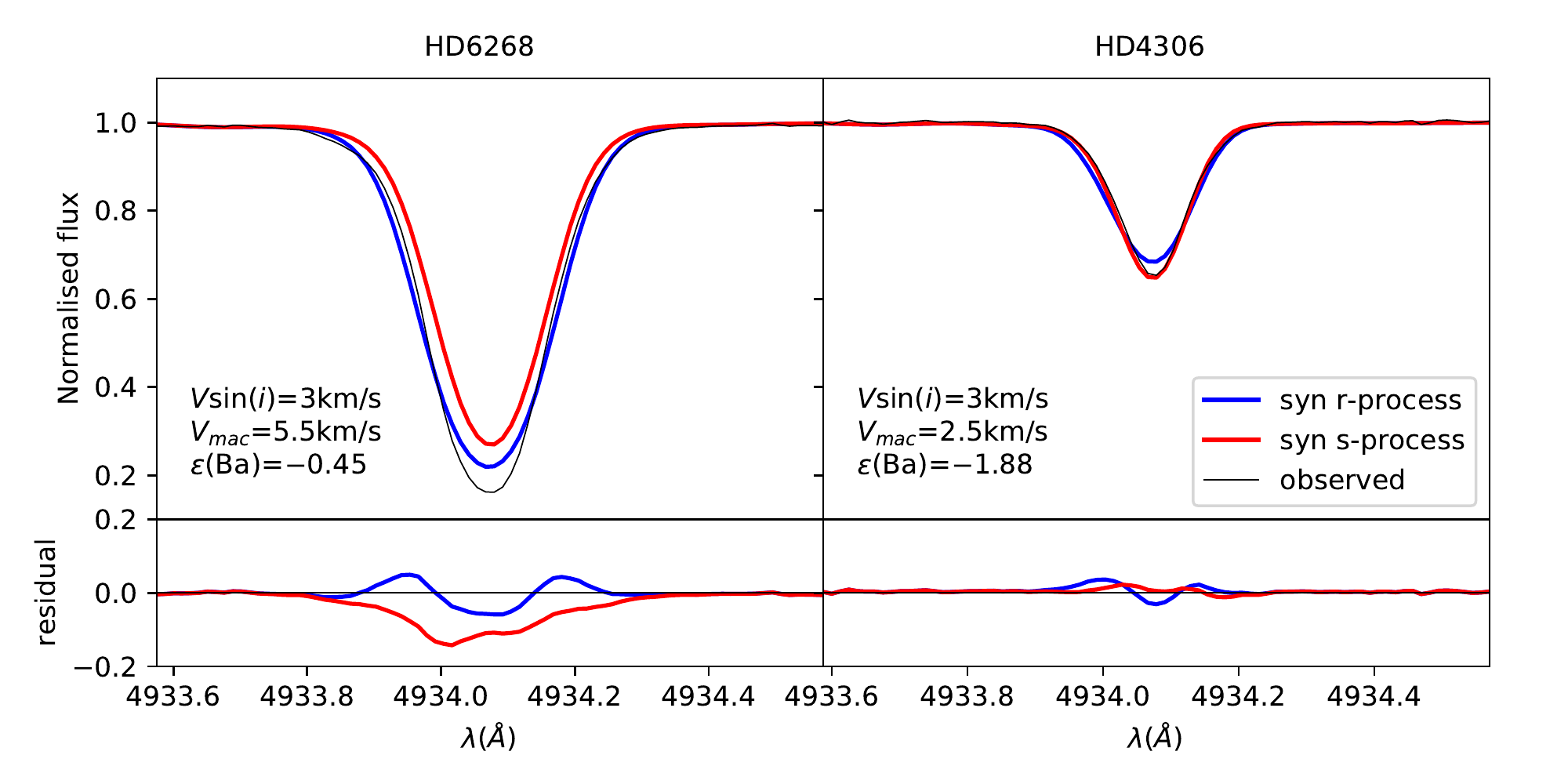}

\caption{Same as Fig. \ref{f4554}, but for Ba II line at 4934\,\AA.}
\label{Fig:f4934}

\end{minipage}
\end{figure*}

Figures~\ref{f5853}-\ref{f6141} show comparisons among the observed profiles (black) and the broadened synthetic ones for the lines at 5853.7\AA\,and at 6141.7 \AA. As mentioned above, these lines do not have strong hfs effects, and therefore they do not show any significant difference in their synthetic profiles between the $s-$  and the $r-$ cases, which appear as a single purple line. The comparison between observed and synthetic lines for 5853.7\AA\,and at 6141.7 \AA\, is good, although the line at 5853.7\AA\, for HD\,4306 is really weak;
these two lines were also used to derive the Ba\,II abundances given in Table\,\ref{tab:ele}. 
On the other hand, the two lines at  4554.0, 4934.1\AA\, are sensitive to hfs effects and therefore can be used to infer the role of $s-$  and $r-$processes in the two stars.
In Figs. \ref{f4554} and \ref{Fig:f4934}, we show the comparison of the observed profiles (black) to the broadened synthetic ones computed for pure r$-$(blue) and s$-$process (red).
In the case of HD\,4306 the observed profiles are very well matched by the synthetic
ones computed assuming a pure $s-$process (see also the small residuals plotted at the bottom of the figures). 
The observed profiles of the 4554.0 and 4934.1\,\AA~lines in HD\,6268 are not very well reproduced by the synthetic ones, either assuming  a pure $s-$ or pure $r-$process. On the other hand, the discrepancies are smaller in the case of the pure  $r-$process (see also the residuals). In HD\,6268, both these two lines are much stronger than the corresponding ones in HD\,4306, thus the mismatch of the total equivalent width between observation and synthetic profile can be ascribed to a problem with the adopted $\xi$ or with deviation from LTE and can be explained only in part by differences in the isotopic composition. An increase of the Ba abundance of 0.7 and 0.5 dex is needed to match the observed equivalent width, assuming an s- and r-process isotopic composition, respectively. The necessary correction is smaller for the r-process case, and the resulting line profile fit is better than that of the s-process assumption. This gives additional support for the conclusion that the barium of HD 6268 most likely has an  r-process origin. In conclusion, by looking at Figs.\,\ref{f4554} and \ref{Fig:f4934} we can say that while the Ba\,II lines in HD\,4306 are fully consistent with a pure $s-$process origin, those in HD\,6268 are better reproduced assuming the predominance of the $r-$process.

\section{Comparison of observed spectra} \label{subsec:comp}

\begin{figure*}
\begin{minipage}{180mm}
\includegraphics[width=165mm]{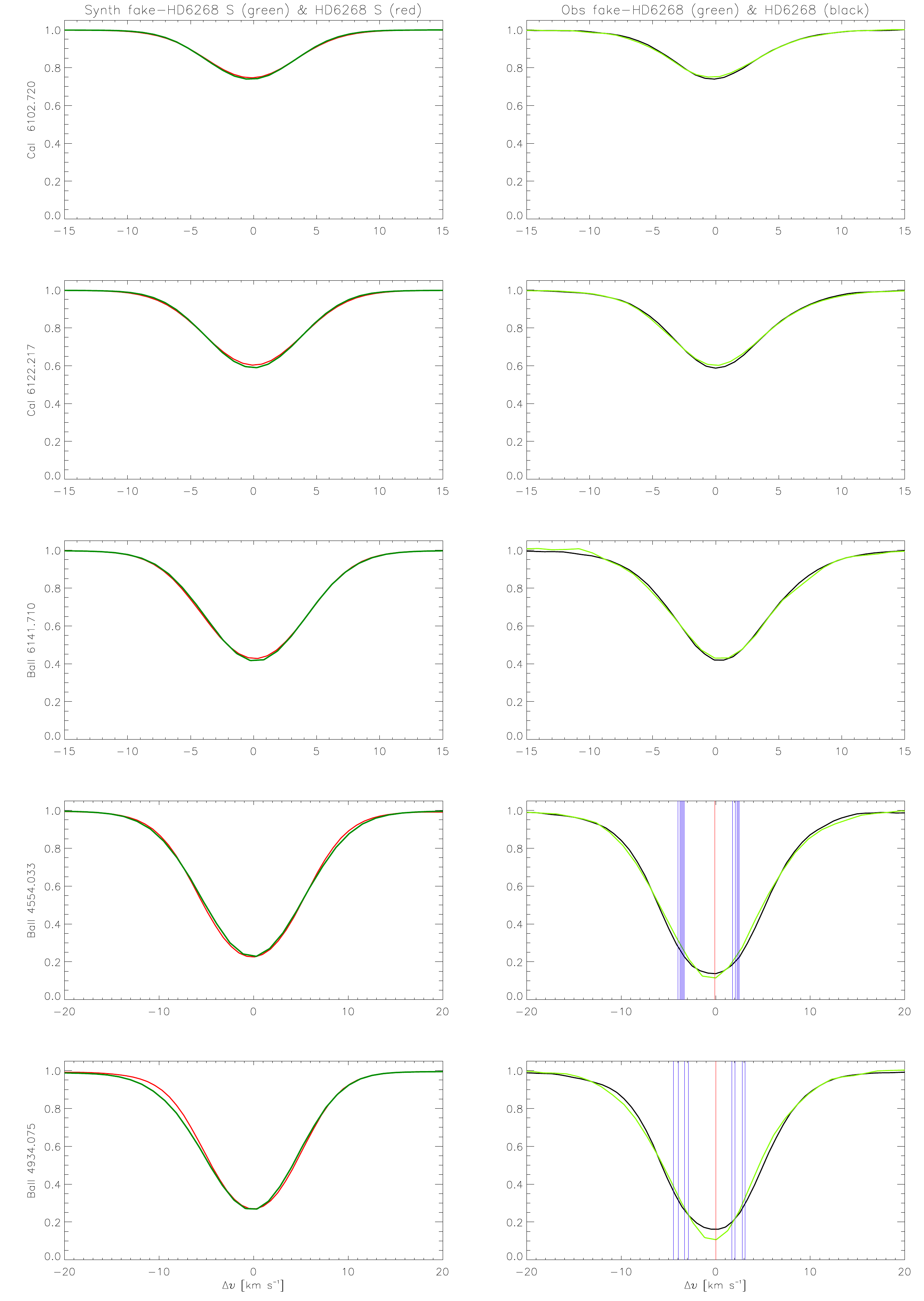}
\caption{Left panels: Comparison of  synthetic pure-s profiles of a scaled fake-HD\,6268 star (dark green) with those of HD\,6268 (red). Right panels:  Comparison of scaled `observed' profiles of a fake-HD\,6268 (light green) with the observed ones of  HD\,6268 (black) (see text). Vertical blue  lines show  the hfs lines of odd Ba isotopes, and vertical red  lines  show lines of the even Ba isotopes. }
\label{Fig:comp}
\end{minipage}
\end{figure*}

The results presented thus far are indeed model dependent. In particular, they depend heavily on the accuracy of the hfs data used to build our synthetic spectra and on the reliability of the spectrum and atmosphere model codes in reproducing the real spectra and atmosphere of the two stars. 
For this reason, it would be worthwhile to perform a relative analysis exclusively using the two observed spectra.
Unfortunately, such an approach would require the two stars to be perfectly equal, apart from the relative contributions of the $s-$ and $r-$process to the formation of Ba. Even if the 4554.0 and 4934.1\,\AA~Ba line profiles depend on the Ba isotope ratios, their profiles are mainly determined by the values of the stellar atmospheric parameters, by the Ba abundance, and by the $\xi$,  $V_{\rm mac}$, and  V$\sin i$ velocities.
In the following, we use the pure-s synthetic spectra of the two stars in a differential way in order to transform the observed spectrum of HD\,4306 into the spectrum this star would have if its atmospheric parameters and velocities were exactly the same as those of HD\,6268. In such a way, our results are still built with the help of the synthetic spectra, but they should be less model dependent than those described in Sect.~\ref{subsec:synth}. 
The idea is to build a spectrum of a pure-s Ba star with the same stellar atmosphere parameters and abundances of HD\,6268 (hereafter fake-HD\,6268) starting from the spectrum of HD\,4306 and to compare it with the spectrum of the real HD\,6268 in order to search for a possible contribution of non-s-processed Ba in the latter star. The lines in HD\,4306 and in the fake-HD\,6268 spectrum will differ in strength (equivalent width) and broadening due to the different stellar atmosphere characteristics and chemical compositions. 
If the line profiles are mainly Doppler profiles, it is easy to scale those of HD\,4306 (in a velocity scale centred at the nominal wavelength of each line) to obtain the broadening of the fake-HD6268 by taking into account the different effective temperatures and macroturbulence velocities in the two stars. On the other hand, if other broadening mechanisms are present (such as, for example, collision broadening, van der Waals broadening, and Stark broadening) it is not easy to take them into account within a simple equation; therefore we adopted a parametric multiplication factor, b, to address their effects.
Thus, we initially changed from the wavelength scale to a velocity scale; that is,  $\Delta\,v=c\times\frac{(\lambda-\lambda_0)}{\lambda_0}$.
Then,  to take into account the differences in broadening, we scaled each pure-s synthetic HD\,4306 line profile to build the corresponding line of the fake-HD\,6268 using the following equation:
\begin{equation}
    \Delta\,v^{\rm \scriptscriptstyle fake-HD\,6268}=b \cdot \Delta\,v^{\rm  \scriptscriptstyle HD\,4306} \sqrt{\frac{T_{\rm eff}^{\rm  \scriptscriptstyle HD6268}}{T_{\rm eff}^{\rm \scriptscriptstyle HD4306}}}\frac{V_{\rm mac}^{\rm \scriptscriptstyle HD6268}}{V_{\rm mac}^{\rm\scriptscriptstyle HD4306}}
.\end{equation}
All the resulting fake-HD\,6268 lines were then scaled to have the same equivalent width of the corresponding HD\,6268 lines 
removing the effects of the different abundances and microturbulences. 
The b value for each line was empirically derived by applying the above procedure to the synthetic pure-s spectra of the two stars  by iterating and  varying {\it b} until we achieved a good match between the synthetic fake-HD\,6268 and the pure-s synthetic HD\,6268 lines. We did not use the observed spectra, since we wanted to avoid incorporating any broadening in the empirically derived b values caused by hfs effects produced by odd isotopes in the Ba lines. In the left panels of Fig.\,\ref{Fig:comp},
we plot the obtained synthetic line profiles of the fake-HD\,6268 (dark green lines) superimposed on the original pure-s synthetic profiles of HD\,6268 (red lines) after re-sampling all of them at steps of 0.1\,km~s$^{-1}$ for all the lines analysed in Sect.\,\ref{subsec:synth}.  This was done with the exception of the Fe\,II line that is blended with the CrI line (4556.163\AA) and the BaII 5853.7\AA~ line that is too week in HD\,4306. 
The values of the b factor are around one (i.e. 0.97, 0.97, 1.15, 1.35, and 1.27 for CaI 6102.7, CaI 6122.2, BaII 6141.7, BaII 4554.0, and BaII 4934.1\,\AA, respectively),  as  expected since we already corrected for the Doppler effects. As can be seen in the left column panels of Fig.\,\ref{Fig:comp}, the fake synthetic profiles match very well those of HD\,6268 apart for the blue wing of the BaII 4934.1\,\AA~ line that has a  known blend with a weak FeI line \citep[see Fig 5 in ][]{Gallagher10} since the two stars have different iron abundances. 
Thus, our procedure works effectively to remove practically all the differences 
between the two stars and  to transform the synthetic pure-s profiles of  HD\,4306  in those we should expect if  HD\,4306 were actually identical to HD\,6268.

Eventually, we apply the so derived individual b factors to scale the observed HD\,4306 lines to obtain the observed fake-HD\,6268 lines by assuming a pure s-process for HD\,6268. In the right column panels of   Fig.\,\ref{Fig:comp},  we plot the observed fake-HD\,6268  profiles (light green lines) superimposed on the corresponding observed profiles of HD\,6268 (black lines)  in order to check our assumption.  
Our procedure removed practically all the differences 
between the two stars and  transform the observed profiles of  HD\,4306  in those we should expect if  HD\,4306 was actually identical to HD\,6268 when we look at the two CaI lines and at the BaII 6141.7 line, which are unaffected by the hypothesis that both stars are pure-s Ba objects. On the other hand, when we apply the same technique  to the 4554.0 and 4934.1\,\AA~ BaII lines, whose profiles are sensitive to the relative contributions of $s-$ and $r-$processes in the two stars, the  observed fake-HD\,6268 profiles do not reproduce the observed profiles of HD\,6268  as well as in the case of the corresponding synthetic ones (left panels in the bottom two lines of Fig.\,\ref{Fig:comp}). In particular, we notice main differences in the cores where the even Ba isotopes shape the lines and in the regions of the hfs lines of odd Ba isotopes, indicated by vertical red  and blue lines in  Fig.\,\ref{Fig:comp}, respectively. So, these results suggest that the assumption  of a pure $s-$process for HD\,6268 should be rejected since it leads to  fake observed profiles of HD\,6268 that are different from the real observed ones, in particular in the region sensitive to s- and r-process contribution.  

We cannot firmly conclude that the barium isotopes in HD\,6268 have a pure r-process pattern. On the other hand, the differences shown in the right panels of the two bottom lines should be due to an effect that does not occur in the other lines. In this respect, the most likely reason is the different pattern of barium isotopes between the two stars; therefore,  these differences can be ascribed to the enrichment of an $r-$process with a large fraction of odd isotopes in HD\,6268, absent from the fake-HD 6268 spectrum. This outcome arises from the differential analysis of synthetic and observed spectra, and so it is almost   independent of the theoretical hfs Ba data; the only dependence is in the negligible hfs effects of the scarce odd isotopes within the s-process pattern.

\section{Conclusions}

  According to our results, the two stars analysed here present two
  different isotopic compositions of barium.
  In fact, the results obtained in Sections~\ref{subsec:synth} and \ref{subsec:comp}
are both consistent with the hypothesis that the barium in HD\,4306 was produced
exclusively by an s$-$process, while the one  in HD\,6268 also shows a contribution  by an $r-$process.
  The predictions of the
  previous chemical evolution models in \citet{Cescutti14} pointed out exactly this
  possibility, that in the Galactic halo both an $s-$process and an $r-$process
  can pollute  the ISM and therefore preserve this differential
  signature. The high [Ba/Fe] stars that typically show a solar or slightly sub-solar [Sr/Ba] 
  ratio are most likely polluted by an $r-$process source, in this case HD 6268 with 
  [Ba/Fe]=0.09 and [Sr/Ba]=0.03.
  On the other hand, an $s-$process source should have polluted the stars with a 
  low [Ba/Fe] and with a [Sr/Ba] that, according to the
  theoretical yields for rotating massive stars, can go from about solar
  (possibly even below solar) up to a [Sr/Ba]$\sim$1.5 \citep{Frisch15,LC18}, in our case HD 4306 with [Ba/Fe]=$-$1.05 and [Sr/Ba]=0.78.
  Therefore, it could be that the light elements' primary production, empirically introduced by \citet{Trava04}, is
  an $s-$process that is activated almost as a primary process by rotation.  This possibility was not excluded by the
  study of \citet{Montes07}, together with the more commonly invoked weak $r-$process. 
  Clearly,  the fact that this happens for two stars does not provide firm conclusions, and more stars must be measured to determine the real statistical impact of the $s-$process in massive stars at low metallicity.
Nevertheless, we think that this work provides insight into the impact of 
 detailed studies of isotopes and the possibilities that high-quality spectra can provide.

\begin{acknowledgements} 
MV and NC acknowledge support by Deutsche Forschungsgemeinschaft (DFG, German Research Foundation) -- project-IDs: 428473034, and 138713538 (SFB~881 ``The Milky Way System'', subproject A04).
CK acknowledges funding from the UK Science and Technology Facility Council (STFC) through grant ST/M000958/1 \& ST/V000632/1.
This work was partially supported by the European Union (ChETEC-INFRA, project no. 101008324 and ChETEC, CA16117).

\end{acknowledgements}

\bibliographystyle{aa}
\bibliography{spectro}

\begin{thebibliography}{56}
\expandafter\ifx\csname natexlab\endcsname\relax\def\natexlab#1{#1}\fi

\bibitem[{{Arcones} \& {Montes}(2011)}]{ArcoMonte11}
{Arcones}, A. \& {Montes}, F. 2011, \apj, 731, 5

\bibitem[{{Arentsen} {et~al.}(2019){Arentsen}, {Prugniel}, {Gonneau},
  {Lan{\c{c}}on}, {Trager}, {Peletier}, {Lyubenova}, {Chen}, {Falc{\'o}n
  Barroso}, {S{\'a}nchez Bl{\'a}zquez}, \& {Vazdekis}}]{2019AA...627A.138A}
{Arentsen}, A., {Prugniel}, P., {Gonneau}, A., {et~al.} 2019, \aap, 627, A138

\bibitem[{{Arlandini} {et~al.}(1999{\natexlab{a}}){Arlandini}, {K{\"a}ppeler},
  {Wisshak}, {Gallino}, {Lugaro}, {Busso}, \& {Straniero}}]{Arla99}
{Arlandini}, C., {K{\"a}ppeler}, F., {Wisshak}, K., {et~al.}
  1999{\natexlab{a}}, \apj, 525, 886

\bibitem[{{Arlandini} {et~al.}(1999{\natexlab{b}}){Arlandini}, {K{\"a}ppeler},
  {Wisshak}, {Gallino}, {Lugaro}, {Busso}, \& {Straniero}}]{arlandini99}
{Arlandini}, C., {K{\"a}ppeler}, F., {Wisshak}, K., {et~al.}
  1999{\natexlab{b}}, \apj, 525, 886

\bibitem[{{Barbuy} {et~al.}(1985){Barbuy}, {Spite}, \&
  {Spite}}]{1985AA...144..343B}
{Barbuy}, B., {Spite}, F., \& {Spite}, M. 1985, \aap, 144, 343

\bibitem[{{Barklem} {et~al.}(2005){Barklem}, {Christlieb}, {Beers}, {Hill},
  {Bessell}, {Holmberg}, {Marsteller}, {Rossi}, {Zickgraf}, \&
  {Reimers}}]{BAR05}
{Barklem}, P.~S., {Christlieb}, N., {Beers}, T.~C., {et~al.} 2005, \aap, 439,
  129

\bibitem[{{Castelli}(2005)}]{castelli05}
{Castelli}, F. 2005, Memorie della Societa Astronomica Italiana Supplementi, 8,
  34

\bibitem[{{Cenarro} {et~al.}(2007){Cenarro}, {Peletier},
  {S{\'a}nchez-Bl{\'a}zquez}, {Selam}, {Toloba}, {Cardiel},
  {Falc{\'o}n-Barroso}, {Gorgas}, {Jim{\'e}nez-Vicente}, \&
  {Vazdekis}}]{2007MNRAS.374..664C}
{Cenarro}, A.~J., {Peletier}, R.~F., {S{\'a}nchez-Bl{\'a}zquez}, P., {et~al.}
  2007, \mnras, 374, 664

\bibitem[{{Cescutti} \& {Chiappini}(2014)}]{Cescutti14}
{Cescutti}, G. \& {Chiappini}, C. 2014, \aap, 565, A51

\bibitem[{{Cescutti} {et~al.}(2013){Cescutti}, {Chiappini}, {Hirschi},
  {Meynet}, \& {Frischknecht}}]{Cescutti13}
{Cescutti}, G., {Chiappini}, C., {Hirschi}, R., {Meynet}, G., \&
  {Frischknecht}, U. 2013, \aap, 553, A51

\bibitem[{{Chiappini} {et~al.}(2008){Chiappini}, {Ekstr{\"o}m}, {Meynet},
  {Hirschi}, {Maeder}, \& {Charbonnel}}]{Chiappini08}
{Chiappini}, C., {Ekstr{\"o}m}, S., {Meynet}, G., {et~al.} 2008, \aap, 479, L9

\bibitem[{{Chiappini} {et~al.}(2011){Chiappini}, {Frischknecht}, {Meynet},
  {Hirschi}, {Barbuy}, {Pignatari}, {Decressin}, \& {Maeder}}]{Chiappini11}
{Chiappini}, C., {Frischknecht}, U., {Meynet}, G., {et~al.} 2011, \nat, 472,
  454

\bibitem[{{Chiappini} {et~al.}(2006){Chiappini}, {Hirschi}, {Meynet},
  {Ekstr{\"o}m}, {Maeder}, \& {Matteucci}}]{Chiappini06}
{Chiappini}, C., {Hirschi}, R., {Meynet}, G., {et~al.} 2006, \aap, 449, L27

\bibitem[{{Choplin} {et~al.}(2018){Choplin}, {Hirschi}, {Meynet},
  {Ekstr{\"o}m}, {Chiappini}, \& {Laird}}]{Choplin18}
{Choplin}, A., {Hirschi}, R., {Meynet}, G., {et~al.} 2018, \aap, 618, A133

\bibitem[{{Collet} {et~al.}(2009){Collet}, {Asplund}, \& {Nissen}}]{Collet09}
{Collet}, R., {Asplund}, M., \& {Nissen}, P.~E. 2009, \pasa, 26, 330

\bibitem[{{Dekker} {et~al.}(2000){Dekker}, {D'Odorico}, {Kaufer}, {Delabre}, \&
  {Kotzlowski}}]{Dekker00}
{Dekker}, H., {D'Odorico}, S., {Kaufer}, A., {Delabre}, B., \& {Kotzlowski}, H.
  2000, in Society of Photo-Optical Instrumentation Engineers (SPIE) Conference
  Series, Vol. 4008, Optical and IR Telescope Instrumentation and Detectors,
  ed. M.~{Iye} \& A.~F. {Moorwood}, 534--545

\bibitem[{{Fran{\c c}ois} {et~al.}(2007){Fran{\c c}ois}, {Depagne}, {Hill},
  {Spite}, {Spite}, {Plez}, {Beers}, {Andersen}, {James}, {Barbuy}, {Cayrel},
  {Bonifacio}, {Molaro}, {Nordstr{\"o}m}, \& {Primas}}]{Franc07}
{Fran{\c c}ois}, P., {Depagne}, E., {Hill}, V., {et~al.} 2007, \aap, 476, 935

\bibitem[{{Francois}(1996)}]{1996AA...313..229F}
{Francois}, P. 1996, \aap, 313, 229

\bibitem[{{Frischknecht} {et~al.}(2016){Frischknecht}, {Hirschi}, {Pignatari},
  {Maeder}, {Meynet}, {Chiappini}, {Thielemann}, {Rauscher}, {Georgy}, \&
  {Ekstr{\"o}m}}]{Frisch15}
{Frischknecht}, U., {Hirschi}, R., {Pignatari}, M., {et~al.} 2016, \mnras, 456,
  1803

\bibitem[{{Gallagher} {et~al.}(2015){Gallagher}, {Ludwig}, {Ryan}, \&
  {Aoki}}]{Gallagher15}
{Gallagher}, A.~J., {Ludwig}, H.~G., {Ryan}, S.~G., \& {Aoki}, W. 2015, \aap,
  579, A94

\bibitem[{{Gallagher} {et~al.}(2010){Gallagher}, {Ryan}, {Garc{\'{\i}}a
  P{\'e}rez}, \& {Aoki}}]{Gallagher10}
{Gallagher}, A.~J., {Ryan}, S.~G., {Garc{\'{\i}}a P{\'e}rez}, A.~E., \& {Aoki},
  W. 2010, \aap, 523, A24

\bibitem[{{Gallagher} {et~al.}(2012){Gallagher}, {Ryan}, {Hosford},
  {Garc{\'{\i}}a P{\'e}rez}, {Aoki}, \& {Honda}}]{Gallagher12}
{Gallagher}, A.~J., {Ryan}, S.~G., {Hosford}, A., {et~al.} 2012, \aap, 538,
  A118

\bibitem[{{Gratton}(1989)}]{1989AA...208..171G}
{Gratton}, R.~G. 1989, \aap, 208, 171

\bibitem[{{Gray}(2008)}]{gray08}
{Gray}, D.~F. 2008, {The Observation and Analysis of Stellar Photospheres}

\bibitem[{{Grevesse} {et~al.}(2007){Grevesse}, {Asplund}, \&
  {Sauval}}]{Grevesse2007}
{Grevesse}, N., {Asplund}, M., \& {Sauval}, A.~J. 2007, \ssr, 130, 105

\bibitem[{{Honda} {et~al.}(2004){Honda}, {Aoki}, {Kajino}, {Ando}, {Beers},
  {Izumiura}, {Sadakane}, \& {Takada-Hidai}}]{honda04}
{Honda}, S., {Aoki}, W., {Kajino}, T., {et~al.} 2004, \apj, 607, 474

\bibitem[{{Jablonka} {et~al.}(2015){Jablonka}, {North}, {Mashonkina}, {Hill},
  {Revaz}, {Shetrone}, {Starkenburg}, {Irwin}, {Tolstoy}, {Battaglia}, {Venn},
  {Helmi}, {Primas}, \& {Fran{\c{c}}ois}}]{Jablonka15}
{Jablonka}, P., {North}, P., {Mashonkina}, L., {et~al.} 2015, \aap, 583, A67

\bibitem[{{Kobayashi} {et~al.}(2020){Kobayashi}, {Karakas}, \&
  {Lugaro}}]{Kobayashi20}
{Kobayashi}, C., {Karakas}, A.~I., \& {Lugaro}, M. 2020, \apj, 900, 179

\bibitem[{{Korotin} {et~al.}(2015){Korotin}, {Andrievsky}, {Hansen}, {Caffau},
  {Bonifacio}, {Spite}, {Spite}, \& {Fran{\c{c}}ois}}]{Korotin15}
{Korotin}, S.~A., {Andrievsky}, S.~M., {Hansen}, C.~J., {et~al.} 2015, \aap,
  581, A70

\bibitem[{{Kurucz}(2005)}]{kurucz05}
{Kurucz}, R.~L. 2005, Memorie della Societa Astronomica Italiana Supplementi,
  8, 14

\bibitem[{{Lambert} \& {Allende Prieto}(2002)}]{LambertAllendePrieto02}
{Lambert}, D.~L. \& {Allende Prieto}, C. 2002, \mnras, 335, 325

\bibitem[{{Limongi} \& {Chieffi}(2018)}]{LC18}
{Limongi}, M. \& {Chieffi}, A. 2018, \apjs, 237, 13

\bibitem[{{Luck} \& {Bond}(1981)}]{1981ApJ...244..919L}
{Luck}, R.~E. \& {Bond}, H.~E. 1981, \apj, 244, 919

\bibitem[{{Luck} \& {Bond}(1985)}]{1985ApJ...292..559L}
{Luck}, R.~E. \& {Bond}, H.~E. 1985, \apj, 292, 559

\bibitem[{{Magain}(1995)}]{Magain95}
{Magain}, P. 1995, \aap, 297, 686

\bibitem[{{Mashonkina} {et~al.}(2017){Mashonkina}, {Jablonka}, {Sitnova},
  {Pakhomov}, \& {North}}]{mashonkina17}
{Mashonkina}, L., {Jablonka}, P., {Sitnova}, T., {Pakhomov}, Y., \& {North}, P.
  2017, \aap, 608, A89

\bibitem[{{McDonald} {et~al.}(2017){McDonald}, {Zijlstra}, \&
  {Watson}}]{2017MNRAS.471..770M}
{McDonald}, I., {Zijlstra}, A.~A., \& {Watson}, R.~A. 2017, \mnras, 471, 770

\bibitem[{{McWilliam}(1998)}]{mcwilliam98}
{McWilliam}, A. 1998, \aj, 115, 1640

\bibitem[{{McWilliam} {et~al.}(1995){McWilliam}, {Preston}, {Sneden}, \&
  {Searle}}]{1995AJ....109.2757M}
{McWilliam}, A., {Preston}, G.~W., {Sneden}, C., \& {Searle}, L. 1995, \aj,
  109, 2757

\bibitem[{{Mel{\'e}ndez} \& {Barbuy}(2002)}]{2002ApJ...575..474M}
{Mel{\'e}ndez}, J. \& {Barbuy}, B. 2002, \apj, 575, 474

\bibitem[{{Montes} {et~al.}(2007){Montes}, {Beers}, {Cowan}, {Elliot},
  {Farouqi}, {Gallino}, {Heil}, {Kratz}, {Pfeiffer}, {Pignatari}, \&
  {Schatz}}]{Montes07}
{Montes}, F., {Beers}, T.~C., {Cowan}, J., {et~al.} 2007, \apj, 671, 1685

\bibitem[{{Pilachowski} {et~al.}(1996){Pilachowski}, {Sneden}, \&
  {Kraft}}]{1996AJ....111.1689P}
{Pilachowski}, C.~A., {Sneden}, C., \& {Kraft}, R.~P. 1996, \aj, 111, 1689

\bibitem[{{Prantzos}(2012)}]{Prantzos12}
{Prantzos}, N. 2012, \aap, 542, A67

\bibitem[{{Prantzos} {et~al.}(2018){Prantzos}, {Abia}, {Limongi}, {Chieffi}, \&
  {Cristallo}}]{Prantzos18}
{Prantzos}, N., {Abia}, C., {Limongi}, M., {Chieffi}, A., \& {Cristallo}, S.
  2018, \mnras, 476, 3432

\bibitem[{{Prugniel} {et~al.}(2011){Prugniel}, {Vauglin}, \&
  {Koleva}}]{2011AA...531A.165P}
{Prugniel}, P., {Vauglin}, I., \& {Koleva}, M. 2011, \aap, 531, A165

\bibitem[{{Raiteri} {et~al.}(1992){Raiteri}, {Gallino}, \& {Busso}}]{Raiteri92}
{Raiteri}, C.~M., {Gallino}, R., \& {Busso}, M. 1992, \apj, 387, 263

\bibitem[{{Rizzuti} {et~al.}(2019){Rizzuti}, {Cescutti}, {Matteucci},
  {Chieffi}, {Hirschi}, \& {Limongi}}]{Rizzuti19}
{Rizzuti}, F., {Cescutti}, G., {Matteucci}, F., {et~al.} 2019, \mnras, 489,
  5244

\bibitem[{{Rizzuti} {et~al.}(2021){Rizzuti}, {Cescutti}, {Matteucci},
  {Chieffi}, {Hirschi}, {Limongi}, \& {Saro}}]{Rizzuti21}
{Rizzuti}, F., {Cescutti}, G., {Matteucci}, F., {et~al.} 2021, \mnras, 502,
  2495

\bibitem[{{Roederer} {et~al.}(2014){Roederer}, {Preston}, {Thompson},
  {Shectman}, {Sneden}, {Burley}, \& {Kelson}}]{Roederer14}
{Roederer}, I.~U., {Preston}, G.~W., {Thompson}, I.~B., {et~al.} 2014, \aj,
  147, 136

\bibitem[{{Rutten}(1978)}]{Rutten78}
{Rutten}, R.~J. 1978, \solphys, 56, 237

\bibitem[{{Sharma} {et~al.}(2016){Sharma}, {Prugniel}, \&
  {Singh}}]{2016AA...585A..64S}
{Sharma}, K., {Prugniel}, P., \& {Singh}, H.~P. 2016, \aap, 585, A64

\bibitem[{{Spite} {et~al.}(2018){Spite}, {Spite}, {Barbuy}, {Bonifacio},
  {Caffau}, \& {Fran{\c{c}}ois}}]{Spite18}
{Spite}, F., {Spite}, M., {Barbuy}, B., {et~al.} 2018, \aap, 611, A30

\bibitem[{{Stoehr} {et~al.}(2007){Stoehr}, {Fraquelli}, {Kamp}, {Kimball},
  {Levay}, {Rogers}, {Smith}, {Thompson}, {White}, \& {Haase}}]{Stoehr07}
{Stoehr}, F., {Fraquelli}, D., {Kamp}, I., {et~al.} 2007, Space Telescope
  European Coordinating Facility Newsletter, 42, 4

\bibitem[{{Thielemann} {et~al.}(2011){Thielemann}, {Arcones}, {K{\"a}ppeli},
  {Liebend{\"o}rfer}, {Rauscher}, {Winteler}, {Fr{\"o}hlich}, {Dillmann},
  {Fischer}, {Martinez-Pinedo}, {Langanke}, {Farouqi}, {Kratz}, {Panov}, \&
  {Korneev}}]{Thielemann11}
{Thielemann}, F.-K., {Arcones}, A., {K{\"a}ppeli}, R., {et~al.} 2011, Progress
  in Particle and Nuclear Physics, 66, 346

\bibitem[{{Travaglio} {et~al.}(2004){Travaglio}, {Gallino}, {Arnone}, {Cowan},
  {Jordan}, \& {Sneden}}]{Trava04}
{Travaglio}, C., {Gallino}, R., {Arnone}, E., {et~al.} 2004, \apj, 601, 864

\bibitem[{{Wu} {et~al.}(2015){Wu}, {Wang}, {Shi}, {Zhao}, \&
  {Grupp}}]{2015AA...579A...8W}
{Wu}, X., {Wang}, L., {Shi}, J., {Zhao}, G., \& {Grupp}, F. 2015, \aap, 579, A8

\end{thebibliography}

\clearpage

\begin{appendix}
\section{Additional Table}

\begin{table}[h!]
\begin{minipage}{160mm}

\caption{List of the atmosphere parameters of HD 6268 from literature. Column 'r' indicates the data type:  'H' is high-resolution spectra,  'L' is low-resolution spectra, and 'P' is photometry. Column 'c' indicates whether the data are original measurements ('O') or a compilation of re-calibrated sources ('C'). Column 'm' indicates the method:   'A' stands for model atmosphere and 'T' stands for empirical calibration.}\label{lit6268}
\begin{tabular}{|r|r|r|r|r|r|l|l|l|l|}
\hline
  \multicolumn{1}{|c|}{Teff} &
  \multicolumn{1}{c|}{err} &
  \multicolumn{1}{c|}{logg} &
  \multicolumn{1}{c|}{err} &
  \multicolumn{1}{c|}{[Fe/H]} &
  \multicolumn{1}{c|}{err} &
  \multicolumn{1}{c|}{r} &
  \multicolumn{1}{c|}{c} &
  \multicolumn{1}{c|}{m} &
  \multicolumn{1}{c|}{Reference} \\
\hline
  4667 &  & 1.25 &  & -2.42 &  & H & O & A & \citet{1981ApJ...244..919L}\\
  4755 &  & 1.2 &  & -2.1 &  & H & O & A & \citet{1985ApJ...292..559L}\\
  4582 &  & 0.75 &  & -2.4 &  & H & O & A & \citet{1985AA...144..343B}\\
  4800 &  100 & 0.84 & 0.03 & -2.56 &  0.05 & H & O & A & \citet{1989AA...208..171G}\\
  4670 &  & 0.75 &  & -2.58 &  & H & O & A & \citet{1995AJ....109.2757M}\\
   4800 & 100 & 0.84 & 0.03 & -2.52 & 0.05 & H & O & A & \citet{1996AA...313..229F}\\
  4700 &  & 1.6 &  & -2.36 &  & H & O & A & \citet{1996AJ....111.1689P}\\
 4705 &  & 1.5 &  & -2.35 &  & H & O & A & \citet{2002ApJ...575..474M}\\
   4600 &  & 1. &  & -2.62 & 0.11  & H  & O & A  & \citet{honda04}\\
  4740 &  & 1.2 &  & -2.32 &  &   & C &   & \citet{2007MNRAS.374..664C}\\
  4735 & 113 & 1.42 & 0.26 & -2.36 & 0.11 & L & O & T & \citet{2011AA...531A.165P}\\
  4570 & 34 & 0.7 & 0.16 & -2.82 & 0.15 & H & O & A & \citet{Roederer14}\\
  4726 & &1.14 & &-2.63 & & H & O & A &\citet{2015AA...579A...8W}\\
    4571 & 161 & 1.13 & 0.39 & -2.63 & 0.18 & L & O & T & \citet{2016AA...585A..64S}\\
  4837 & 125 & 1.59 &  &  &  & P & O & T & \citet{2017MNRAS.471..770M}\\
  4662 & 47 &1.37 & 0.14 &-2.47 & 0.07 & H & O & A & \citet{2019AA...627A.138A}\\
 
\hline
\end{tabular}
\end{minipage}
\end{table}

\end{appendix}

\end{document}